\documentclass[reprint,amsmath,amssymb,superscriptaddress,prb]{revtex4-2}
\usepackage[rgb]{xcolor}
\usepackage{graphicx}
\usepackage{booktabs}
\usepackage{hyperref}
\newcommand{\be}{\begin{equation}}
\newcommand{\ee}{\end{equation}}
\newcommand{\beqn}{\begin{eqnarray}}
\newcommand{\eeqn}{\end{eqnarray}}

\begin{document}

\title{Quantum entanglement in the multicritical disordered Ising model}
\author{Istv\'an A. Kov\'acs}
\affiliation{Department of Physics and Astronomy, Northwestern University, Evanston, Illinois 60208}
\affiliation{Northwestern Institute on Complex Systems, Northwestern University, Evanston, Illinois 60208}
\affiliation{Department of Engineering Sciences and Applied Mathematics, Northwestern University, Evanston, Illinois 60208}
\email{istvan.kovacs@northwestern.edu}

\date{\today}

\begin{abstract}
Quantum entanglement at critical points is often marked by universal characteristics. 
Here, the entanglement entropy is calculated at the quantum multicritical point of the random transverse-field Ising model (RTIM). We use an efficient implementation of the strong disorder renormalization group method in two and three dimensions for two types of disorder. For cubic subsystems we find a universal logarithmic corner contribution to the area law $b\ln(\ell)$ that is independent of the form of disorder. Our results agree qualitatively with those at the quantum critical points of the RTIM, but with new $b$ prefactors due to having both geometric and quantum fluctuations at play. By studying the vicinity of the multicritical point, we demonstrate that the corner contribution serves as an ``entanglement susceptibility'', a useful tool to locate the phase transition and to measure the correlation length critical exponents.
\end{abstract}

\maketitle

\section{Introduction}
Quantum critical points (QCPs) occur in the ground state of quantum systems by tuning a quantum control parameter that governs quantum fluctuations. Quantum multicritical points (QMCPs) emerge at the junction of two or more quantum phase transitions,  resulting in novel universality classes \cite{sachdev}. While QCPs have been well characterized theoretically, our understanding of QMCPs remains much more limited. From an experimental perspective, QMCPs are expected to be less elusive to study than QCPs %, especially in ferromagnetic systems
\cite{buried1,buried2, buried3, buried4}, see for example
%
%In ferromagnetic systems, quantum critical points (QCPs) are often hard to access experimentally as they might either change to first-order transition or get buried inside an intervening phase  \cite{buried1,buried2, buried3, buried4}. 
%
%At the intersection of two phase-transitions, quantum multicritical points 
%As a recent example, 
the recent experimental work on the ferromagnetic QMCP %has been studied experimentally 
in the disordered compound $\mathrm{Nb}_{1-y}\mathrm{F}_{2+y}$  \cite{friedemann}.
%Indeed, QMCPs have been proposed as key concepts to understand the onset of the ordered phase in a range of quantum systems, 
%
On the theoretical side, our recent study showed that the QMCP of the ferromagnetic random transverse-field Ising model (RTIM) exhibits ultraslow, activated dynamic scaling \cite{multi}, governed by an infinite disorder fixed point (IDFP) \cite{fisher, danielreview}. The dominant role of disorder ensures that the applied strong disorder renormalization group (SDRG) method  \cite{mdh,im} is asymptotically exact \cite{danielreview,mccoywu,shankar,young_rieger96,bigpaper},  meaning that the obtained numerical results approach the exact results at large scales. 
In this paper, our goal is to quantify the universal aspects of quantum entanglement at the QMCP of the RTIM. Our results contribute to a better understanding of the universal properties of quantum many body-systems in the vicinity of quantum phase-transitions  \cite{entanglement_review,amico,area,laflorencie}. 
We consider the ground state of the system, $|\varPsi\rangle$, and measure the entanglement between a subsystem, ${A}$ and the rest of the system, ${ B}$, by the von Neumann entropy of the reduced density matrix,
$\rho_{A}={\rm Tr}_{{B}} | \varPsi \rangle \langle \varPsi |$ as
\begin{equation}
{\cal S}_A=-{\rm Tr}_{ A}\left(\rho_{A} \log_2{ \rho_{A}}\right)\;.
\label{eq:S}
\end{equation}
Known as the `area law' \cite{area}, ${\cal S}$ is generally expected to scale with the area of the interface separating ${\cal A}$ and ${\cal B}$ in the ground state. At QCPs, however, there are often additional universal corrections, which can be dominant in one-dimensional systems \cite{holzhey,vidal,Calabrese_Cardy04}. 
In higher dimensions it is much more challenging to study quantum entanglement in interacting systems. %, our understanding about bipartite entanglement is far less complete. %the known results are almost exclusively for two-dimensional ($d=2$) models. Considering non-interacting systems, 
%For free bosons the area law holds even in gapless phases \cite{boson}. On the contrary, for gapless free-fermionic systems with short-range hoppings and a finite Fermi surface there is a logarithmic factor to the area law \cite{fermion}. 
At the QCP of two-dimensional interacting systems there are additional logarithmic terms, which are expected to be universal, as demonstrated 
%
%$d=2$ examples include 
for multiple models, including the transverse-field Ising model  \cite{2d_Ising}, the antiferromagnetic Heisenberg model \cite{2d_Heisenberg}, %For the latter the logarithmic terms are associated to two sources: i) corners on the boundary of the subsystem and ii) non-trivial topology in the bulk.
%For $d=2$ systems described by conformal field theory, such as the square lattice 
and the quantum dimer model \cite{rokhsar_kivelson}. %the log-correction to the area law is also known to be universal and related to corners \cite{2d_conf}. 

%%%%%%%%%%% FIG 1  %%%%%%%%%%%%%%%%%%%%%%%%%%%%%%%
\begin{figure}
%\vskip-7mm
\centering
\includegraphics[width=6.3cm,angle=0]{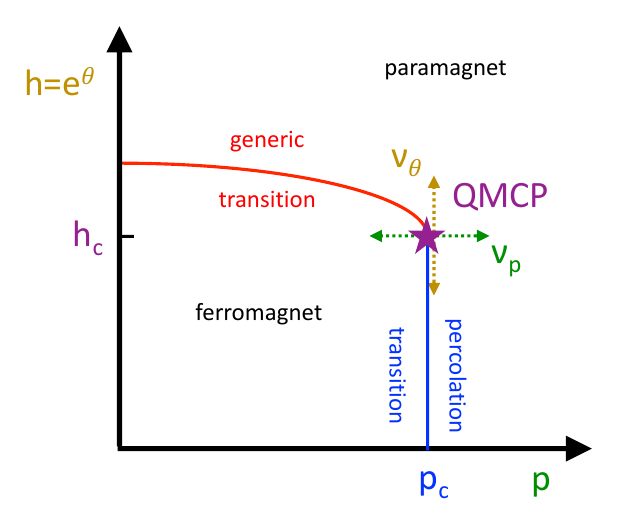}
\vskip -5mm
\caption{ 
\label{fig:RTIM} 
Phase diagram of the RTIM in two and higher dimensions. The QMCP (purple) emerges at the junction of the percolation transition at the bond dilution parameter $p=p_c$ (blue line) and the generic disordered universality class (red), when the $h$ magnetic field is tuned to its critical value. %At the junction of the two transitions the quantum multicritical point emerges (purple, QMCP). 
Deviations from the QMCP are governed by two correlation length exponents, $\nu_\theta$ and $\nu_p$, corresponding to the two control parameters.
}
%\vskip -3mm
\end{figure}
%%%%%%%%%%% FIG 1  %%%%%%%%%%%%%%%%%%%%%%%%%%%%%%%

Disordered systems have been also extensively studied, with the RTIM as a prominent example  \cite{refael}, as at an IDFP disorder fluctuations dominate over quantum fluctuations, simplifying the analytic and numerical treatment  \cite{motrunich00,lin00,karevski01,lin07,yu07,2dRG,ddRG}. 
In addition to critical exponents, the SDRG method also offers an efficient way to calculate the entanglement properties \cite{EPL}. While the area law is again found to be valid in disordered magnets, the total entanglement entropy is not universal and not extremal at the critical point in higher dimensions. Yet, in the RTIM there is a singular, logarithmic corner contribution to the entanglement entropy that is universal and extremal at the critical point, as shown in $d=2$, 3 and 4 \cite{EPL}. 

In this paper, we show that the same kind of scenario holds at the so far uncharted QMCP of the RTIM in two and three dimensions. As our main result, we quantify the logarithmic corner contribution to the entanglement entropy of cubic subsystems with high precision and show that it is universal, i.e., independent of the form of disorder. In addition, we show that just like at the QCP of the RTIM \cite{EPL}, the corner contribution serves as an ``entanglement susceptibility'', determining the location of the QMCP as well as the correlation length critical exponents \cite{ansell}. % and universal aspects of the QMCP.
%This corner contribution can be directly accessed and measured to high precision. 

\section{Model and Methods}

The Hamiltonian of the RTIM can be expressed as
\begin{equation}
{\cal H} =
-\sum_{\langle ij \rangle} J_{ij}\sigma_i^x \sigma_{j}^x-\sum_{i} h_i \sigma_i^z\;,
\label{eq:H}
\end{equation}
where the $\sigma_i^{x,z}$ Pauli-matrices represent spins at sites $i$ of a $d$-dimensional cubic lattice. The spins interact through the $J_{ij}$ nearest neighbor couplings, and are exposed to the $h_i$ transverse fields. Both the couplings and the fields are non-negative random numbers, drawn from some distributions. To test the universality of the results we will use two different types of disorder as in Refs.~\cite{ddRG, EPL, 2dRG, multi}. 
For both types of disorder, the couplings are uniformly distributed in the interval $(0,1]$. The transverse fields are either constant $h_i=h,\, \forall i$ (\emph{fixed-$h$} disorder), or are drawn independently from the interval $(0,h]$ (\emph{box-$h$} disorder). %, whereas for \emph{fixed-$h$} disorder we apply a constant transverse field $h_i=h,\, \forall i$. 
The choice of fixed-$h$ disorder can be motivated by experimental realizations of the model where the transverse field is homogeneous, e.g.~in $\rm{LiHo}_x\rm{Y}_{1-x}\rm{F}_4$ \cite{silevitch}. 
%Note that as far as universal properties are concerned, the shape of the distributions is found to be irrelevant, as long as they are non-singular.

Just like for the QCP, the QMCP of the RTIM is studied with the quantum control parameter given by the logarithmic variable $\theta=\ln(h)$  \cite{2dRG,ddRG,EPL}. 
To arrive at the QMCP, the bond percolation probability $p$ must be tuned to its critical value $p_c$, as illustrated in Fig.~\ref{fig:RTIM}. For sufficiently small fields, we observe a quantum phase transition dictated by the classical percolation transition of the lattice \cite{2qcp, senthil_sachdev}. 
%in higher dimensions, the RTIM also undergoes a second order quantum phase transition by tuning the bond percolation probability $p$ for sufficiently weak external fields . This percolation QCP happens at the classical bond percolation critical point $p_c$, where the system is fragmented into separate clusters. 
%Along this line the ground state of ${\cal H}$ is given by a set of independently ordered GHZ clusters, the magnetic domains, with the same asymptotic geometry as for classical percolation. %The critical exponents are known exactly in $d=2$ and to high precision in $d=3$ as summarized in Table \ref{table:1}. 
This percolation line ends at the QMCP, where it meets the line of the generic QCP transition. Along the generic transition line the critical behavior falls in the same universality class as the undiluted ($p=0$) system \cite{2dRG,ddRG,EPL}.
%For $p>p_c$, at least one giant percolating cluster is present in the system, providing the basis of the generic QCP by tuning the external field $h$ to its critical value (which depends on $p$), see Fig.~\ref{fig:RTIM}. %Along the quantum transition line, a 'quantum' percolation takes place that has been characterized in both 2 and 3 dimensions. 
%
At the QMCP a new universality class emerges, characterized by a new set of critical exponents, due to the interplay of both geometric and quantum fluctuations, see Ref.~\cite{multi} and Table~\ref{table:1}. 
%As the QMCO (and both QCPs) 

The SDRG method offers a very efficient way to obtain the ground state of the RTIM \cite{2dRG,ddRG} by iteratively creating an effective description of the ground state and low-energy excitations. %During the SDRG method  \cite{im} the largest local terms in the Hamiltonian in Eq.~(\ref{eq:H}) are successively eliminated and new Hamiltonians are generated through perturbation calculation.
%During the SDRG procedure, an effective system of the ground state and low-energy excitations of the RTIM is created iteratively, following the successive application of two simple decimation steps. 
At each decimation step of the process the largest local term in the Hamiltonian in Eq.~(\ref{eq:H}) is eliminated. There are two options: the largest term could either be the strongest $J$ coupling or the largest $h$ transverse field in the system. % depending on whether the largest local term is a $J$ coupling or a $h$ transverse field. Then, new terms are generated between remaining spins by second-order perturbation method as follows. 
Second-order perturbation theory then dictates the emergence of new, weak couplings depending on the two options as follows. 
 \emph{$J$-decimation:} when the largest term in the system is a coupling, $J_{ij}$, the two connected spins tend to be aligned at low energies and can be merged into an effective spin cluster of the joint moment, $\tilde{\mu}=\mu_i+\mu_j$. This effective spin is then placed in an effective transverse field, $\tilde{h}=h_i h_j/J_{ij}$. 
 \emph{$h$-decimation:} when the largest term in the system is a transverse field, $h_i$, the spin does not contribute to the magnetic properties of the system at low energies and can be eliminated. However, new weak effective couplings needs to be placed between each pair of neighboring spins, $j$ and $k$, $\tilde{J}_{jk}=J_{ji}J_{ik}/h_i$.
%\end{itemize}
%
In the case when a coupling is generated between a pair of spins that are already interacting by another coupling, the maximum of the two $J$ couplings is taken. This choice is known as the \emph{maximum rule}, which is known to be a valid approximation at an IDFP where the distribution of the couplings becomes extremely broad.
Note that as a result, in all cases, the new effective terms are smaller than the eliminated terms. At each successive step of the SDRG, another spin is eliminated as the energy scale is continuously lowered, until all degrees of freedom have been decimated out.
In practice, the most efficient implementation of the SDRG method works in a parallel manner \cite{ddRG}, relying on graph algorithms to obtain the same results as the above mentioned conceptual picture, but in nearly linear time as a function of the number of spins. 
%
%If at any point two parallel couplings appear between the same sites, the maximum of them is taken. Application of this \emph{maximum rule} is justified at an IDFP, where the distribution of the couplings becomes broader and broader without limits. 
%
%
%While a na\"{\i}ve algorithm generates many new couplings in the system in the $h$-decimation steps, the efficient implementation reduces the number of links in each step, requiring only $t \sim {O}(N \log N)$ CPU time and ${O}(N)$ memory to obtain the ground state a system with $N$ sites and $E\sim dN$ edges in $d$ dimensions. %, see Fig.~\ref{fig_sdrg_fast}.
%
%After decimating all degrees of freedom, 
The ground state of the RTIM is then obtained as a collection of independent ferromagnetic clusters of various sizes -- created at each $h$-decimation step. In each cluster, all spins point in the same directions as all others, known as a GHZ state $\frac{1}{\sqrt{2}}\left(|\uparrow \uparrow  \dots  \uparrow\rangle +
|\downarrow \downarrow \dots  \downarrow\rangle \right)$. 
%The ground state GHZ clusters are illustrated in two dimensions at the generic QCP and QMCP in Fig.~\ref{fig:clusters}. 
%
%This makes the SDRG an ideal tool to study both the generic QCP and the QMCP. It has been shown that the maximum rule also leads to an exact mapping between the SDRG and a graph algorithm, enabling  a highly efficient SDRG implementation \cite{ddRG} that runs in nearly linear time in terms of the number of spins. 

%As we will discuss later in more detail, 
While the emerging clusters are generally fractal-like disconnected objects \cite{EPL,emergence}, each contributes equally to the entanglement entropy of a subsystem, as long as it is intersected by the subsystem in a way that there are some site(s) inside and outside \cite{refael_moore04}. With the definition in Eq.~(\ref{eq:S}) each such intersected cluster contributes to the amount of $\log_2 2=1$, turning the calculation of the entanglement entropy into a cluster counting task.

%The non-trivial, fractal geometry of these clusters is then the key to understand the entanglement patterns of these models.
%Independently from the cluster geometry, each GHZ cluster contributes $\log_2 2=1$ to the entanglement entropy of a single subsystem in Eq.~(\ref{eq:S}) if the cluster has at least one site inside and one site outside of the subsystem, otherwise the contribution is $0$  \cite{refael_moore04}. Thus, calculation of the entanglement entropy for the RTIM is equivalent to a classical cluster counting problem. While at the percolation QCP the clusters look like percolation clusters, at the generic QCP and QMCP the clusters are fundamentally different, as they are disconnected at large scales \cite{EPL,emergence}. Therefore, counting the clusters that are intersecting the boundary of a subsystem not only requires checking the cluster structure around the boundary of the subsystem but also in its entire volume.
%
%
%As the main result of the field, the number of clusters crossed by the contour (surface) of a subsystem is not strictly proportional to the surface area of the subsystem, but there is an additional universal logarithmic singularity at the critical point, known as the corner contribution \cite{EPL}. Such corner contribution depends sensitively on the shape of the subsystem, as shown for the percolation QCP \cite{kovacs, kovacs3dperc}. 
%Multiple geometries, geometric method

%%%%%%%%%%% FIG 1  %%%%%%%%%%%%%%%%%%%%%%%%%%%%%%%
\begin{figure}
%\vskip-7mm
\centering
\includegraphics[width=6.cm,angle=0]{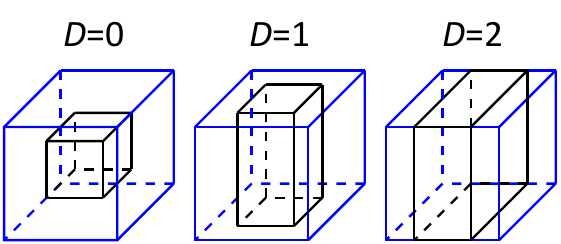}
\vskip -2mm
\caption{ 
\label{fig:shapes} 
Subsystem shapes used in the geometric method in $d=3$ \cite{EPL}. Each subsystem spans the full $L$ length in $D$ directions, with a size of $\ell=L/2$ in the remaining directions. With periodic boundary conditions, the corner contribution is only present for the cubic subsystem, $D=0$. 
}
%\vskip -3mm
\end{figure}
%%%%%%%%%%% FIG 1  %%%%%%%%%%%%%%%%%%%%%%%%%%%%%%%

Interestingly, the entanglement entropy depends sensitively on the shape of the subsystem. As shown at the percolation \cite{kovacs3dperc,ansell} and generic \cite{EPL} QCPs, subsystems with sharp corners lead to universal corner contributions. For example, a cubic subsystem in $d=3$ is expected to yield the critical result of 
\be
{\cal S}^{(3)}(\ell)=a_2 \ell^2 + a_1 \ell + {\cal S}_{\rm cr}^{(3)}\;
%b^{(3)}\ln(\ell)
+O(1)\;
\label{S^d}
\ee
in the limit of large system sizes $1/L\to\infty$ when the $\ell$ linear size of the subsystem is proportional to the system size.
Here ${\cal S}_{\rm cr}^{(3)}=b^{(3)}\ln(\ell)+O(1)$, and only the $b^{(3)}$ prefactor is universal \cite{EPL}, with the values summarized in Table~\ref{table:1}.
Outside the critical point, the finite correlation length is expected to lead to a finite corner contribution, as we will discuss later. %In comparison, outside of the critical point, the contribution of the corners is expected to be at most a size-independent constant, as the correlation length is finite. 
Form this form it is apparent that the corner contribution is relatively small compared to the non-universal terms. Yet, it can be measured directly to high precision using the so called \emph{geometric method} \cite{EPL,ansell}, at least in the case of periodic boundary conditions applied here. The idea is to use additional measurements that have a different shape, fully spanning the system in $D$ dimensions, incorporating a different amount of each term seen for a cubic subsystem due to a different amount of surface elements, like corners, edges and facets. In $d=3$, in addition to cubes, we also consider columns ($D=1$, has edges, but no corners) and slabs ($D=2$, no edges and no corners), as illustrated in Fig.~\ref{fig:shapes}. More generally, in $d$ dimensions, we considered $d$ different geometries with $D=0,1,\dots,d-1$ to obtain the corner contribution \cite{EPL} as
\be
{\cal S}^{(d)}_{\mathrm{cr}}=\sum_{D=0}^{d-1}{\left(-\frac{1}{2}\right)^{D}\binom{d}{D}{\cal S}^{(d)}_{D}}\;.
\label{corners}
\ee
Note that the geometric method cancels out all other terms, not only on average over samples, but exactly in each sample even at small sizes, where there are additional finite-size effects contributing to the asymptotic terms. Hence, the geometric method often provides high-precision results with relatively small finite-size effects.

\section{Results}

%%%%%%%%%%% FIG 1  %%%%%%%%%%%%%%%%%%%%%%%%%%%%%%%
\begin{figure}[ht]
\centering
\includegraphics[width=0.95\linewidth]{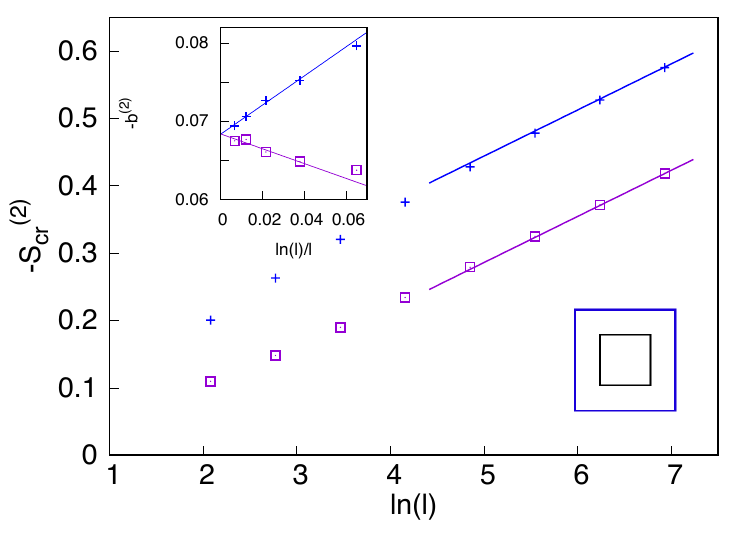}
\includegraphics[width=0.95\linewidth]{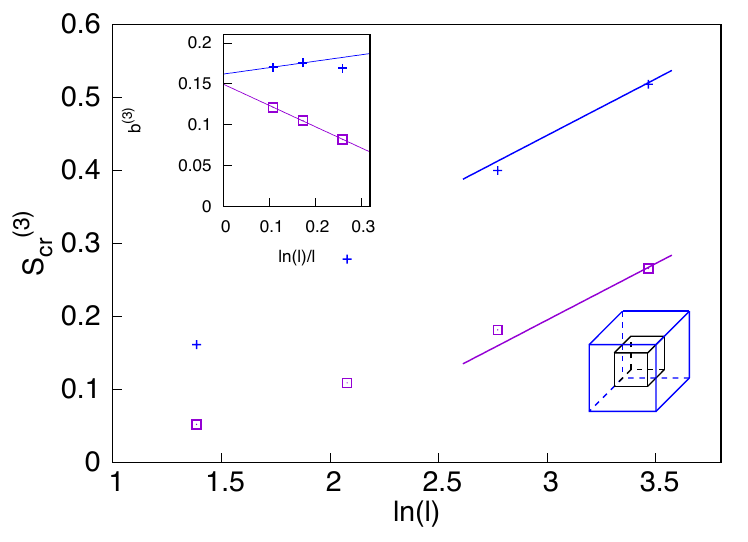}
\vskip -.5cm
\caption{
\label{fig:b}
Corner contribution to the entanglement entropy of cubic subsystems in the $d=2$ (top) and $d=3$ (bottom) models for fixed-$h$ ($+$) and box-$h$ ($\Box$) disorder realizations.
Insets: Extrapolation of the effective prefactors of the logarithm are shown as calculated by two-point fits. As an indication of universality, the extrapolated values are disorder independent as listed in Table~\ref{table:1}. %Extrapolation of $b$ through two-point fits. 
The error of the datapoints is smaller than the size of the symbols.
}
\end{figure}
%%%%%%%%%%% FIG 1  %%%%%%%%%%%%%%%%%%%%%%%%%%%%%%%

%To confirm the universality of the results, we use two different forms of randomness, as used before in Refs.~\cite{ddRG, EPL, 2dRG, multi}. 
%as presented in Table \ref{table:1}. %In addition to the well studied QCPs, recently we have located the quantum multicritical point (Fig.~\ref{fig:RTIM}) and measured the multicritical exponents to high precision in both two and three dimensions. 
%The current best estimates of the exponents are presented in Table \ref{table:1}. 
 %As expected, these $\theta$ values of the QMCP are much smaller than those at the generic QCP ($p=1$), as the correlation clusters need to overcome the dilution of the lattice.
%

The location of the QCPs and QMCP are known to high precision, as listed in Table \ref{table:1}. %To study the QMCP, the bond probability is set to its indicated critical value %$p_c(\mathrm{2})=0.5$ and $p_c(\mathrm{3})=0.248812$ 
%\cite{3dperc}. 
Here, we also list the relevant critical and multicritical exponents, al of which are known to be universal, i.e., randomness independent \cite{multi}. %, with the relevant exponents indicated in Table~\ref{table:1}. %
The known $b$ values of the corner contribution to the entanglement entropy
are also listed here for $d=2,3$ at the percolation and generic QCPs.
We study large systems up to a linear size of $L=2048$ in $d=2$ and $L=64$ in $d=3$. 
The number of realizations used in the numerical calculations at the QMCP is typically $100\,000$, apart from the largest sizes, where we have at least $50\,000$ samples. 
The total computational effort exceeded 10 CPU years.
%to be surpassed in our planned calculations to the extent that is required to sufficiently reduce the statistical error with reasonable run time on Northwestern's state-of-the-art servers.
%and for the two disorder distributions.

We implemented the `geometric method' to obtain the corner contribution as well as the other prefactors $a_i$ in Eq.~(\ref{S^d}). As expected, the area law is found to be valid at the QMCP, with non-universal $a_i$ prefactors. In $d=2$, $a_1=0.237(1)$ for box-$h$ disorder and $a_1=0.662(1)$ for fixed-$h$ disorder. In $d=3$, $a_2=0.163(1)$ and $a_1=-0.11(1)$ for box-$h$ disorder, with $a_2=0.546(1)$ and $a_1=-0.24(1)$ for fixed-$h$ disorder.

%Plot $a_2$ $a_1$. 
At the QMCP, we see clear evidence of a logarithmic corner contribution in both $d=2$ and $d=3$, as shown in Fig.~\ref{fig:b}, with the insets indicating the two-point fits of $b^{(d)}$ from consecutive sizes. 
As a clear sign of universality, the extrapolated $b^{(d)}$ values are found to be disorder independent, and are listed in Table \ref{table:1}. In both two and three dimensions, the $b^{(d)}$ prefactors are between those at the generic and percolation QCPs. 
%As noted in Ref.~\ref{EPL}, it follows from the geometric considerations that the sign of $b^{(d)}$ has to be negative in even dimensions and positive in odd dimensions.

%%%%%%%%%%% FIG 2  %%%%%%%%%%%%%%%%%%%%%%%%%%%%%%%
\begin{figure}
\centering
\includegraphics[width=\linewidth]{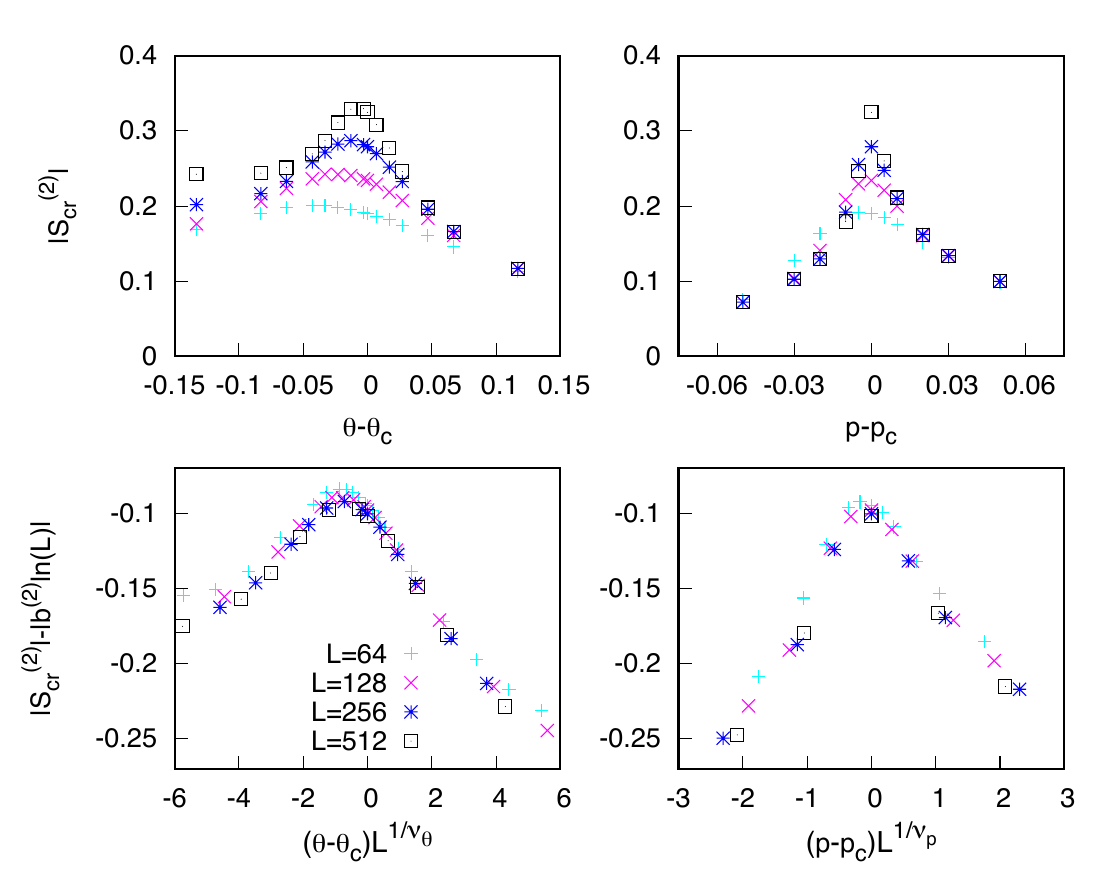}
\vskip -.5cm
\caption{
\label{fig:gr}
Corner contribution to the entanglement entropy in the vicinity of the QMCP for box-$h$ disorder in $d=2$.
Left: Varying $\theta$ at $p_c$ (brown in Fig.~\ref{fig:RTIM}). Right: Varying $p$ at $\theta_c$ (green in Fig.~\ref{fig:RTIM}).
Bottom: Data collapse with the estimated value of $b^{(2)}$ as well as the known values of the $\nu_\theta$ and $\nu_p$ critical exponents, listed in Table \ref{table:1}.
The error of the datapoints is smaller than the size of the symbols.
}
\end{figure}
%%%%%%%%%%% FIG 2  %%%%%%%%%%%%%%%%%%%%%%%%%%%%%%%

%%%%%%%%% REVISED UNTIL HERE
While the entanglement entropy is not extremal at higher-dimensional QCPs or at the QMCP, the corner contribution is only present at the phase transitions, suggesting an extremal ${\cal S}_{\rm cr}^{(d)}$ as a function of either $\delta_\theta=\theta-\theta_c$ or $\delta_p=p-p_c$.
%Off-critical results in both directions:
%Illustration for 2d box-$h$ $L=512$
For $\delta>0$ we arrive at the paramagnetic Griffiths-phase, while $p-p_c<0$ leads to a ferromagnetic Griffiths-phase.
Note that along the $p=p_c$ critical line for $\theta<\theta_c$, we asymptotically expect to see the percolation critical behavior as the Griffiths-phase is only present for $\theta>\theta_c$ in this case.
In Ref.~\cite{multi} it was found that the vicinity of the QMCP is highly anisotropic, as the $\nu$ correlation length critical exponent is different for the two control parameters, as listed in Table~\ref{table:1}.
We have therefore also studied the behavior of the corner contribution to the entanglement entropy outside the critical point and measured
${\cal S}_{\rm cr}^{(d)}(L,\delta)$ as a function of either $\delta_{\theta}$ or $\delta_p$.
%
%In the ordered phase, $\delta<0$, and for $\xi<\ell$ the giant cluster behaves as a so called global cluster, which has $1$ contribution to the entropy for all position, orientation and shape of the subsystem. As shown in the Appendix in odd (even) dimensions, after averaging for all positions a global cluster has a contribution $2^{1-d}$ ($0$) to the corner entropy. Approaching the critical point for $\xi \gtrsim \ell$ these giant clusters have a finite, but $\delta$-dependent contribution, that we omit in the following analysis.
%
In the upper panels of Fig.~\ref{fig:gr} ${\cal S}_{\rm cr}^{(d)}(L,\delta)$ is presented
for box-$h$ disorder in $d=2$ for $10^4$ samples, showing a clear peak at the QMCP in both directions.
%For any $d$ the corner-entropy is expected to be extremal around the critical point and its value
Outside the multicritical point, the corner contribution is limited by the finite correlation length, $\xi \sim |\delta|^{-\nu}$, leading to the substitution  $\ell \to \xi$. %, with $\xi \sim |\delta|^{-\nu}$
%being the correlation length. 
Therefore, close to the multicritical point, in the Griffiths-phase, the corner contribution satisfies the scaling relation
\be
{\cal S}_{\rm cr}^{(d)}(L,\delta)-b^{(d)} \ln L = f(\delta L^{1/\nu}), 
\ee
as illustrated by the data collapse in the lower panels of Fig.~\ref{fig:gr}. Here we have used the known $d=2$ estimates for the $\nu_p$ and $\nu_\theta$
correlation length critical exponents, listed in Table \ref{table:1}. 
These results underline that the corner contribution is not only universal, but provides a systematic way to locate the multicritical points in higher dimensional interacting quantum systems, as well as $b$ and the $\nu_\theta$ and $\nu_p$ critical exponents. Let us emphasize again that the behavior of the corner term is in stark contrast to the full entanglement entropy, which is generally non-universal and non-maximal at the critical point in higher dimensions.
% \cite{2dRG,ddRG}, $\nu=1.24,~0.98$ and $0.78$ in $d=2,3$ and $4$, respectively.
%
%\section{Conclusion}

\begin{table}[ht]
\footnotesize
\caption{
\textbf{Critical and multicritical  properties of the RTIM:} %``$-$" indicates unknown results to be determined in this project. 
 The universal $b$ prefactors of the corner contribution to the entanglement entropy at the QMCP are indicated in bold. 
f stands for fixed-$h$ disorder, while b indicates box-$h$ disorder. The results of this work are indicated in bold.
%CFT stands for conformal field theory.
 \label{table:1}}
 \centering
 \begin{tabular}{cc|c|c|c}  
   && Percolation&Generic& QMCP\\ 
    & & QCP  \cite{kovacs, kovacs3dperc, yu07} & QCP  \cite{EPL}  &  \cite{multi} \\ \hline
  $d=2$  & $p_c$ or $\theta_c$ & $0.5$ bond& $-0.17034(2)$ f & $-0.481(1)$ f \\
     &      & $0.592746$ site & $1.6784(1)$ b& $0.783(1)$ b \\ 
&$\nu_\theta$    & NA & $1.24(2)$ &$1.382(7)$  \\ 
&$\nu_p$    & $4/3\sim1.333$ & NA & $1.168(10)$  \\ 
 &$b^{(2)}$     & $-\frac{5\sqrt{3}}{36\pi}\approx -0.07657$  & $-0.029(1)$ & $\mathbf{-0.0684(4)}$ \\  \hline
  $d=3$  & $p_c$ or $\theta_c$ & $0.248812$ bond& $-0.07627(2)$ f & $-0.5055(10)$ f \\
     &      & $0.311608$ site & $2.5305(10)$ b& $0.770(1)$ b \\ 
     &$\nu_\theta$    & NA  & $0.98(2)$  &$1.123(10)$\\ 
&$\nu_p$    & $0.8762(12)$  & NA &$0.86(1)$\\ 
 &$b^{(3)}$      & $1.72(3)$  & $0.012(2)$ & $\mathbf{0.155(10)}$ \\ 
 %square (contour,3d)    & $-0.158(10)$  & $-$ & $-$ \\ 
 %square (planar,3d)    & $-$  & $-$ & $-$ \\ 
  \end{tabular}
%  \vskip-5mm
  \end{table}
  
\section{Discussion}

We have studied the quantum entanglement properties at the multicritical point (QMCP) of a paradigmatic interacting quantum system (RTIM) in both two and three dimensions. % has been unexplored for multiple reasons, including the 
While the area law is found to be valid for cubic subsystems, we have identified universal logarithmic corner contributions. The results at the QMCP are found to be between that of the two participating critical lines---correspondign to the percolation and generic QCPs---in both $d=2$ and $d=3$.
%The corresponding prefactor $b^{(d)}$ encapsulates universal information from correlations at all orders in the shape of the underlying spin clusters \cite{EPL}.
%
This work contributes to the emerging picture of how universal features of entanglement manifest at higher dimensional QCPs and QMCPs. For a single subsystem, geometric singularities, like corners, play an essential role and lead to a universal prefactor $b$, akin to a critical exponent, which is independent from the usual set of exponents. In contrast to traditional critical exponents $b$ aggregates higher-order correlations \cite{EPL}, and is expected to showcase a non-trivial dependence on the shape of the subsystem.

%Further steps: shape-dependence, skeletal entanglement, multipartite entanglement
%The corner contributions are expected to depend sensitively on the shape of the subsystem, an interesting future direction to explore. 
%We aim to extend these results to several directions. First we mention 
Measuring the shape-dependence of the entanglement entropy at QCPs and at the QMCP is an interesting future direction, also related to recently proposed models of quantum communication \cite{ravi}. 
%in two and three dimensions. As there are no results at the QMCP, we will measure the entanglement for the first time for square and cubic subsystems, as well as the shape-dependence. %Through more general shapes, such as tilted squares and cubes. 
For example, in $d=2$ the shape-dependence can be confronted with the results of conformal invariance. %in $d=2$, we can check if at the QCP and QMCP conformal invariance is possible in these disordered systems in a statistical sense, when averaged over many realizations.
Currently, the most complete results are available at the percolation QCP, where in two dimensions the system is conformally invariant, enabling a full analytic treatment supported by high-precision numerical methods \cite{kovacs,ansell}. %We have shown that each $\gamma_k$ corner has an independent contribution to $b$  \cite{kovacs}.
%, given by:
%\begin{equation}
%b=-\frac{\left.c'(Q)\right|_{Q=1}}{24} \sum_k \left(  \dfrac{\pi}{\gamma_k}-  \dfrac{\gamma_k}{\pi}+ \dfrac{\pi}{2 \pi-\gamma_k}-\dfrac{2 \pi-\gamma_k}{\pi}\right)\;,
%\label{cardy_peschel}
%\end{equation}
%where $\gamma_k$ is the interior angle at each corner, and $c(Q)$ is the central charge of the $Q$-state Potts model, with $c'(1)=\dfrac{5 \sqrt{3}}{4 \pi}$ corresponding to percolation, see Fig.~\ref{fig:perc}. %and the two sets of terms come from the interior and exterior contribution.
%From this it is clear that while the value of $b$ depends on the shape of the subsystem, different subsystem shapes provide no additional information compared to a square at the two-dimensional percolation QCP. %This result has been confirmed extensively by studying various subsystem shapes, such as triangles, tilted squares, etc, as shown in Fig.~\ref{fig:shape}. 
Detailed shape-dependence of cluster counts have been also obtained numerically for the percolation QCP in three dimensions \cite{3dperc,ansell}. % where we no longer have conformal predictions. 
In general, especially in the lack of conformal invariance, the shape-dependence of the corner contributions is expected to be universal but non-trivial, meaning that different subsystem shapes might extract different information on the entanglement patterns. %, as illustrated in Fig.~\ref{fig:perc}.
%Studying subsystems of various shapes is valuable tool that can reveal new information for different shapes and check if the predictions of conformal invariance hold. While for the percolation QCP in $d=2$ any subsystem shape reveals the same information, this is no longer expected to be the case in other systems or in $d=3$. 
As the simplest possibility, line segments of length $\ell$ are of special interest \cite{ansell}. Line segments are special cases of \emph{skeletal entanglement}, where the subsystem is a zero-measure volume of the full system, % \cite{skeletal}. While still in its infancy, skeletal entanglement is a very interesting new concept that has 
offering additional universal results \cite{skeletal}. %, with a lot of new physics yet uncharted.

Another key question that arises is whether studying multipartite entanglement can provide further insights \cite{szalay,zimbi}. %In other words, do we gain qualitatively new, universal insights about critical quantum systems by quantifying multipartite entanglement? Recently, 
As shown recently in the one-dimensional RTIM \cite{jay}, the multipartite entanglement structure \cite{calabrese} is qualitatively different in otherwise similar disordered quantum chains \cite{connect}. % but also universal if we consider the appropriate geometric configurations.
The RTIM results also showed that in the appropriate geometric scaling limit, %by considering the right geometric configurations, 
multipartite entanglement measures are universal and provide deeper information than bipartite entanglement. 
%Our recent results showcased universal quantum correlation between two subsystems at any distance in $d=1$, when two subsystems of size $\ell$ are separated at a distance $r=\alpha\ell$. 
On the contrary to the entanglement entropy, where only the (leading order of the) corner contribution is universal, in the case of both the entanglement negativity and mutual information, %${\cal E}$ and ${\cal I}$, 
the entire multipartite measure was found to be universal \cite{jay}. 
%, at least in the appropriate (linear) scaling limit.
Extending these results to non-adjacent subsystems in higher dimensional QCPs and QMCPs is an exciting future direction. Our results can be also extended to the RTIM with long-range interactions \cite{long-range,long-rangeCP,long-range3d}, motivated by materials like $\rm{LiHo}_x\rm{Y}_{1-x}\rm{F}_4$ \cite{silevitch}.

\section*{Acknowledgments}
%double degree--degree distance distribution
This work was supported by the National Science Foundation under Grant No.~PHY-2310706 of the QIS program in the Division of Physics.

%\bibliography{tomography-sources}

\end{document}